\documentclass{webofc}
\usepackage[varg]{txfonts}   
\usepackage{physics}
\usepackage{setspace}

\newcommand{\xt}{{\mathbf{x}}}
\newcommand{\yt}{{\mathbf{y}}}
\newcommand{\bt}{{\mathbf{b}}}
\newcommand{\rt}{{\mathbf{r}}}
\newcommand{\ud}{\, \text{d}}

\newcommand{\ampli}{{\mathcal{N}}}

\newcommand{\qs}{Q_{\text{s}}}
\begin{document}
\title{Hot spots in a proton}
%
%

\author{\firstname{Sami} \lastname{Demirci}\inst{1,2}\fnsep
\and
\firstname{Tuomas} \lastname{Lappi}\inst{1,2}\fnsep\thanks{\email{tuomas.v.v.lappi@jyu.fi}} \and
        \firstname{Sören} \lastname{Schlichting}\inst{3}\fnsep
}

\institute{Department of Physics, P.O.~Box 35, 40014 University of Jyv\"{a}skyl\"{a}, Finland
\and
           Helsinki Institute of Physics, P.O.~Box 64, 00014 University of Helsinki, Finland
\and
Fakult\"at für Physik, Universit\"at Bielefeld, D-33615 Bielefeld, Germany
          }

\abstract{%
We explore the consequences of gluonic hot spots inside the proton for the initial eccentricities in a proton-nucleus collision, and the constraints on the parameters describing these hot spots from coherent and incoherent exclusive vector meson production cross sections in deep inelastic scattering.
We show that geometric fluctuations of hot spots inside the proton are the dominant source of eccentricity whereas color charge fluctuations only give a negligible correction.
We find that the coherent cross section is sensitive to both the size of the target and the structure of the probe. The incoherent cross section is dominated by color fluctuations at small transverse momentum transfer ($t$), by proton and hot spot sizes as well as the structure of the probe at medium $-t$ and again by color fluctuations at large $-t$. 

}
\maketitle
\section{Introduction}
\label{intro}

It has become increasingly clear that the internal geometrical structure of the nucleon plays an important role in both heavy ion collisions and the initial stages of quark-gluon plasma formation, as well as observables that can be measured in new deep inelastic scattering experiments. Very commonly this internal structure is understood in terms of gluonic ``hot spots'' in the nucleon. Some recent examples of such models include the energy dependent hot spot model of Ref.~\cite{Cepila:2017nef}, the IPglasma model for the initial stages of heavy ion collisions~\cite{Mantysaari:2016ykx,Mantysaari:2016jaz},and  the hot spot model of Ref.~\cite{Albacete:2016pmp} used to understand the ``hollowness''  of the proton as measured in elastic proton-proton scattering. Often the hot spots are related to a quark model approach~\cite{Traini:2018hxd,Dumitru:2020gla}, but they have also been speculated to form a self-similar cascade of hot spots within hot spots~\cite{Kumar:2021zbn}, or simply parametrized and studied within Bayesian fits of flow observables in heavy ion collisions~\cite{Moreland:2018gsh}.

Our purpose here is to develop a simple  model for gluonic hot spots within a proton. Rather than putting in all the bells and whistles to fit a maximum amount of data, we want our model to be analytically tractable in order to develop a better understanding of the physical mechanisms involved. We have used this model to understand the initial eccentricity in proton-nucleus collisions~\cite{Demirci:2021kya} (see also \cite{Demirci:2023ejg}), and studied how the hot spot properties can be constrained by exclusive vector meson production data from deep inelastic scattering~\cite{Demirci:2022wuy}.

\section{Simple model for hot spots}

Our model for the nucleon enables us to analytically average over two kinds of degrees of freedom. There is an average over color configurations in the spirit of the McLerran-Venugopalan model, with color charges generating a field around them regularized by an infrared regulator scale $m$. Separately, we average over the locations of $N_q$ color charge hot spots with a Gaussian profile of size $r_H$, themselves distributed with a Gaussian density inside a proton of radius $R$. 
In equations, this averaging procedure can be expressed as color charges with a two point function given by a hot spot density profile
\begin{equation}
\expval{\rho^a(\xt) \rho^b(\yt)}_{c} = \sum_{i=1}^{N_q} \mu^2 \left(
\frac{\xt+\yt}{2} - \bt_i \right) \delta^{(2)}(\xt-\yt) \delta^{ab},
\quad
 \mu^2(\xt)=\frac{\mu^2_0}{2\pi r_H^2} \exp
\left[ -\frac{\xt^2}{2r_H^2} \right]
\end{equation}
and hot spots averaged over the size of the proton\footnote{Note that we fix the center of mass of the hot spot system at the origin of our coordinate system.}
\begin{equation}
 \langle \langle \mathcal{O} \rangle
\rangle = \left( \frac{2 \pi R^2}{N_q} \right) \int \prod^{N_q}_{i=1}
\Big[ \ud^2\bt_i T(\bt_i-\mathbf{B})\Big] 
 \delta^{(2)}\left(
\frac{1}{N_q}\sum _{i=1}^{N_q}\bt_i - \mathbf{B}\right)  \langle
\mathcal{O} \rangle_{c}
\quad
 T(\bt)=\frac{1}{2\pi
R^2}\exp \left[ -\frac{\bt^2}{2R^2} \right]. 
\end{equation}
From the color charge density we can calculate the color field and the  Wilson line in the usual way, and from these obtain the initial glasma field in a heavy ion collision, or the scattering amplitude for exclusive vector meson production. In this work we calculate these observables to lowest order in $\rho$ to enable an analytical averaging over degrees of freedom.

\section{Eccentricity}
 A hydrodynamical calculation of flow observables requires density profiles of the initial stage, including the effects of the hot spot structure. We thus calculate the eccentricity in our hot spot model. We use the analytically calculable energy density at $\tau=0$ (at the expense of introducing an additional UV regulator ~\cite{Lappi:2006hq}). In our theorist's idealized proton-nucleus collision the proton has both color charge and hot spot fluctuations, but is treated as linear in the color charge density. The nucleus, on the other hand, is taken to be infinite in size and smooth, with only color charge fluctuations, but treated fully to all orders in the CGC field.


\begin{figure}[tbh!]
\includegraphics[height=2.4cm]{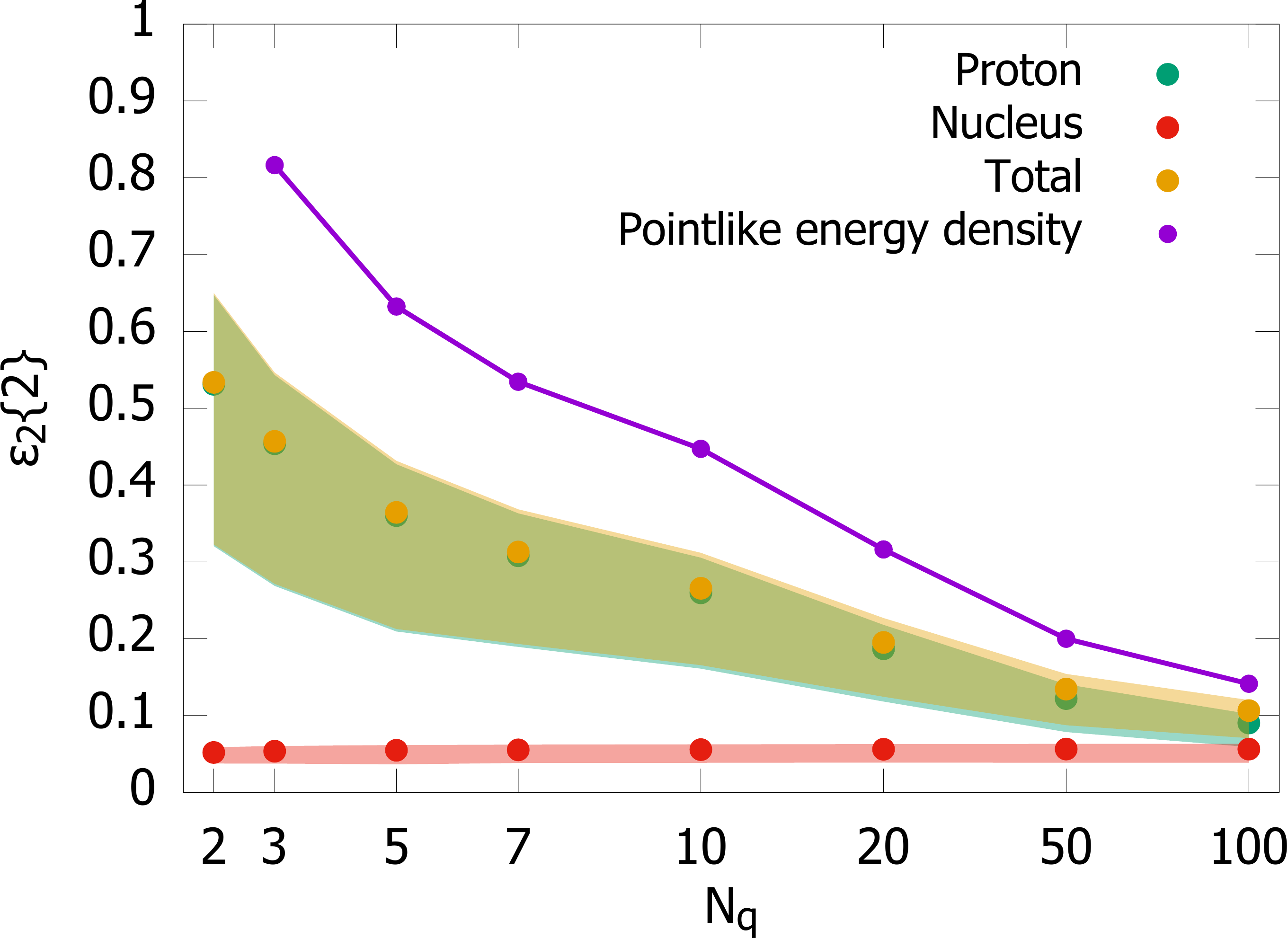}
\includegraphics[height=2.4cm]{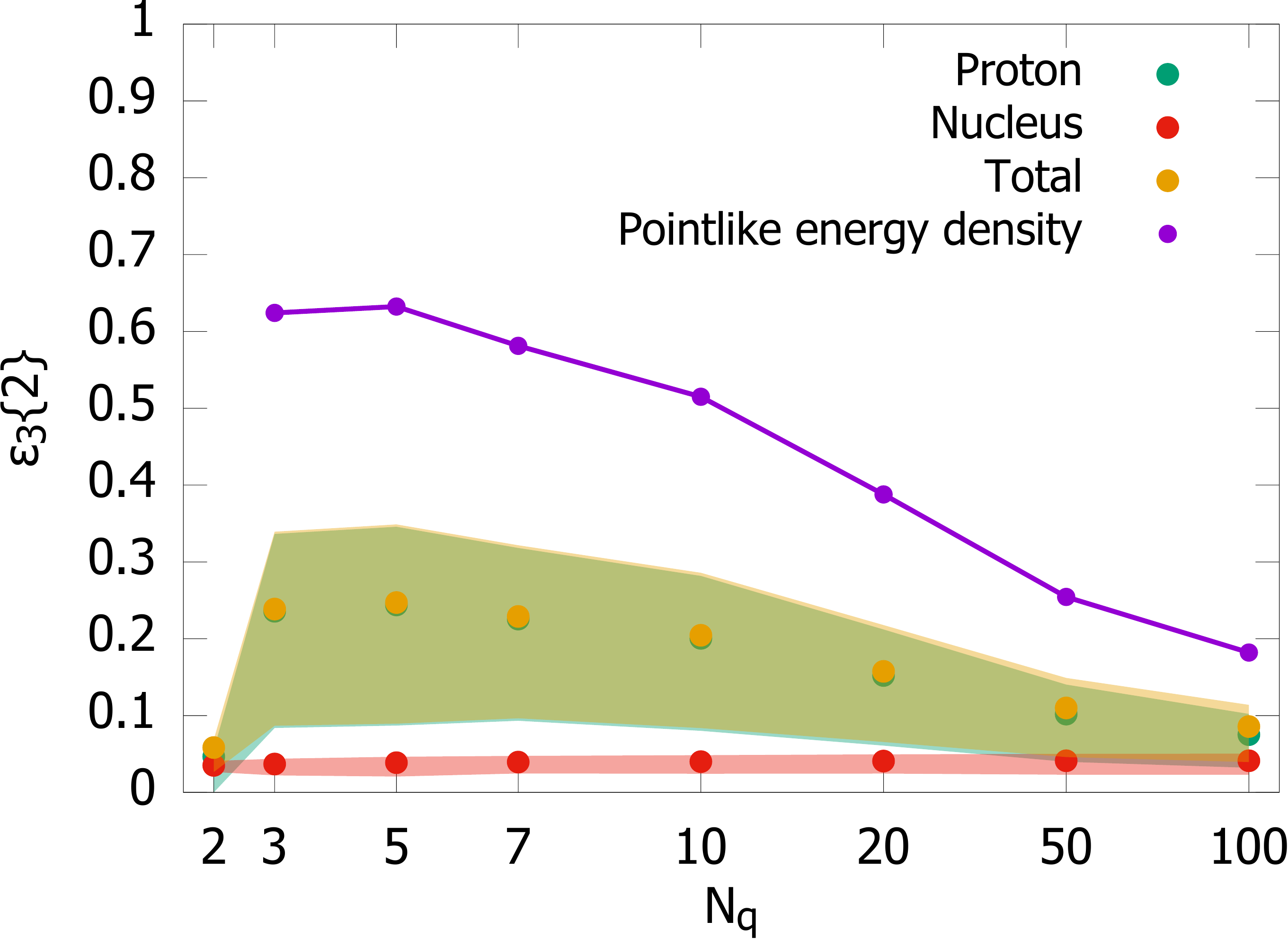}
\includegraphics[height=2.4cm]{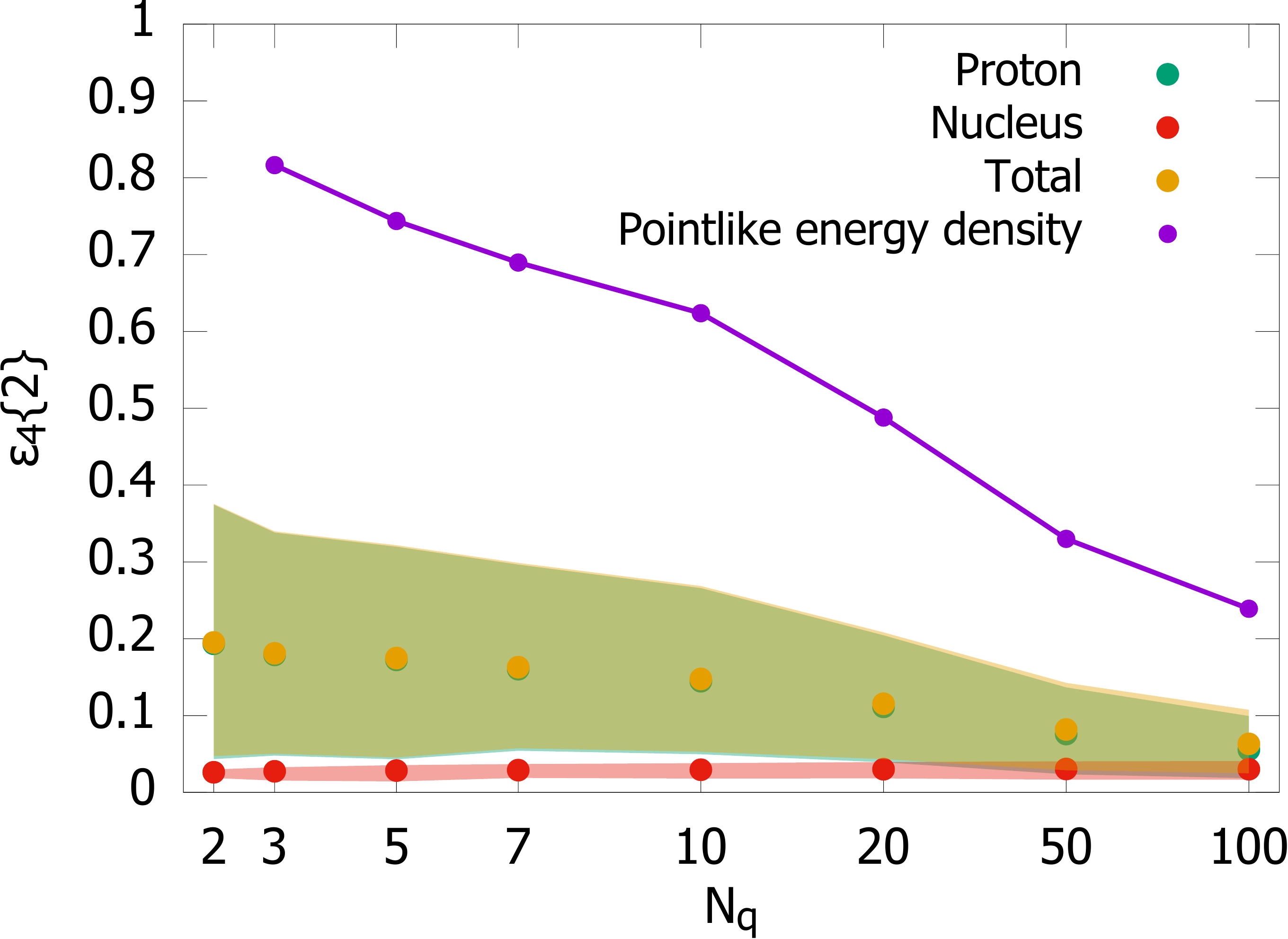}
\includegraphics[height=2.4cm]{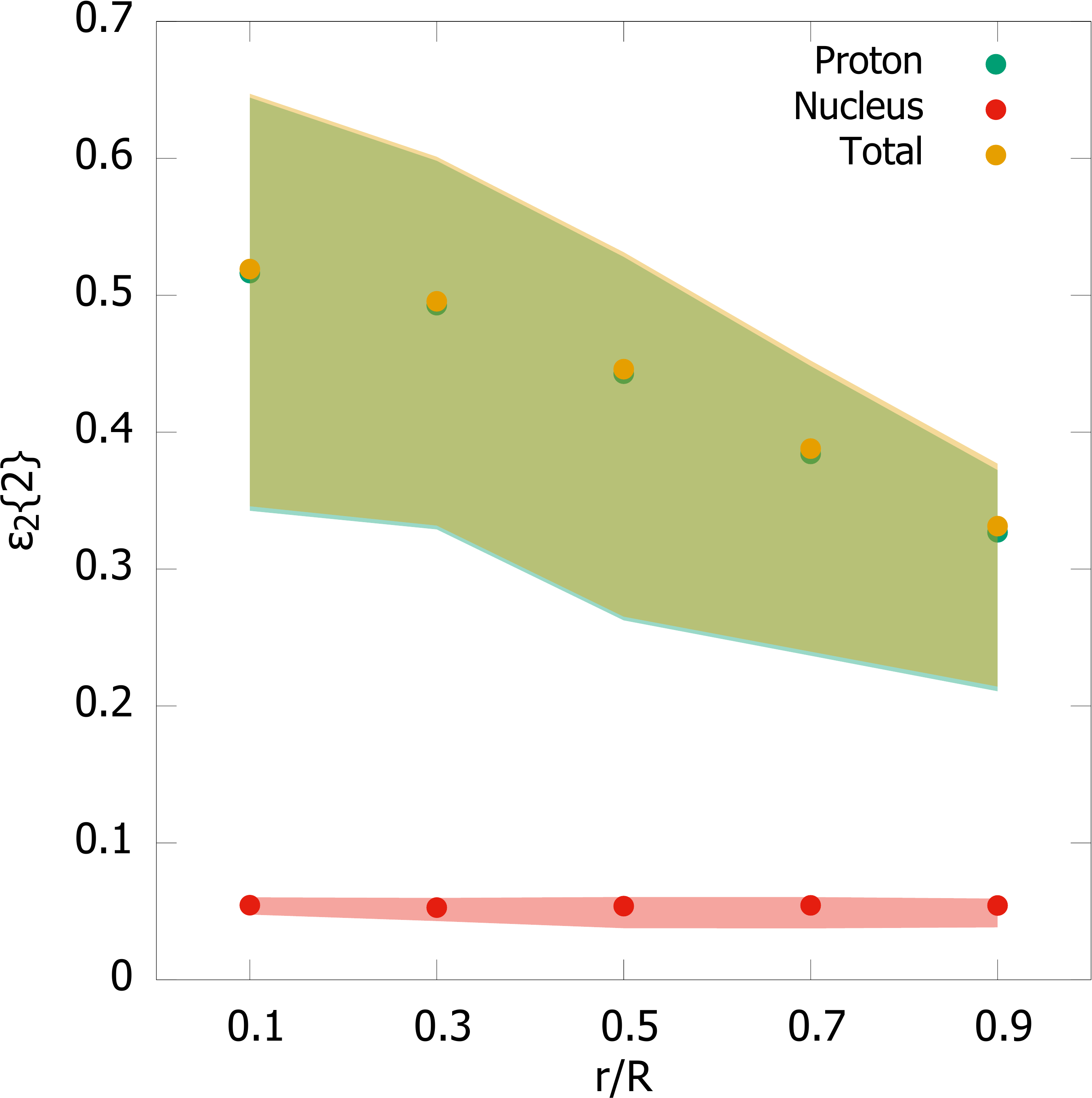}
\caption{
The eccentricities for the second, third and fourth harmonics $\varepsilon_n{2}, n=2,3,4$ as a function of the number of hot spots. The different symbols show the contribution of the nucleus-side charge fluctuations and the proton-side hot spot and charge fluctuations. Also shown is a limiting case of pointlike hotspots. The plot on the right shows the dependence on the hot spot size for $N_q=3$. The error bars correspond to varying the IR regulator $m$ by $\pm 50\%$.
}\label{fig:eps}
\end{figure}

The eccentricities
\begin{equation}
\left(\varepsilon_n\{2\} \right)^2= \frac{\int \ud^2\xt\ud^2\yt~|\xt|^n~|\yt|^n~ 
e^{in(\theta_{\xt}-\theta_{\yt})}~\left< \varepsilon(\xt)\varepsilon(\yt)\right>}{
\int \ud^2\xt\ud^2\yt~|\xt|^n~|\yt|^n~ 
\left< \varepsilon(\xt)\varepsilon(\yt)\right>
}
\end{equation}
are determined by the two-point correlation function of the energy density. 
Inspecting results for the eccentricities shown in fig.~\ref{fig:eps}, we can see that  the proton hot spots dominate the correlation structure except for very large $N_q$. Although the energy density correlator is strongly dependent on the UV cutoff, this does not affect the eccentricities, which are quite sensitive to the IR regulator for the Coulomb tail of the color field $m$, and to the hot spot size $r_H$.

\section{Exclusive vector mesons}
A natural next question to ask is whether the parameters of such a hot spot model could be constrained by exclusive vector meson production. Here, a useful framework is provided by the Good-Walker picture, where different kinds of quantum fluctuations in the target hadron or nucleus, such as nucleons in a nucleus, hot spots in a nucleon, color charges as in the MV model, or different size color dipoles, can be related to the elastic and incoherent cross sections. In an elastic process the outgoing target is in the same state  as the incoming one, in a quasielastic process fluctuations to some limited set of final states are allowed, and the incoherent cross section is given by the difference of the two:

\begin{equation}
\sigma_{\text{el}} \sim \abs{\bra{\Omega} \ampli \ket{\Omega}}^2
\quad 
\sigma_{\text{quasiel}} \sim \sum_{\Omega'} \abs{\bra{\Omega'} \ampli \ket{\Omega}}^2
\quad
\sigma_{\text{incoh}} \sim \sum_{\Omega'} \abs{\bra{\Omega'} \ampli \ket{\Omega}}^2
- \abs{\bra{\Omega} \ampli \ket{\Omega}}^2,
\end{equation}
where $\ket{\Omega}$ is the incoming state and $\ket{\Omega'}$ states to which fluctuations are allowed. The elastic vector meson production amplitude becomes particularly simple in the nonrelativistic limit for the meson light cone wavefunction,
$
\Psi_{V}(z,|\rt|)\propto \delta \left(z-\frac{1}{2}\right),
$
where in  this limit the only dependence on $\rt$, the size of the quark-antiquark dipole, comes in through the virtual photon wavefunction. In our hot spot model, we can evaluate both the coherent and incoherent cross sections, in particular the target averages, almost completely analytically, apart from some of the integrals over the dipole size. In particular, we do not assume a factorizable impact parameter profile. 

\begin{figure}
\resizebox{\textwidth}{!}{
\includegraphics[height=1.03cm]{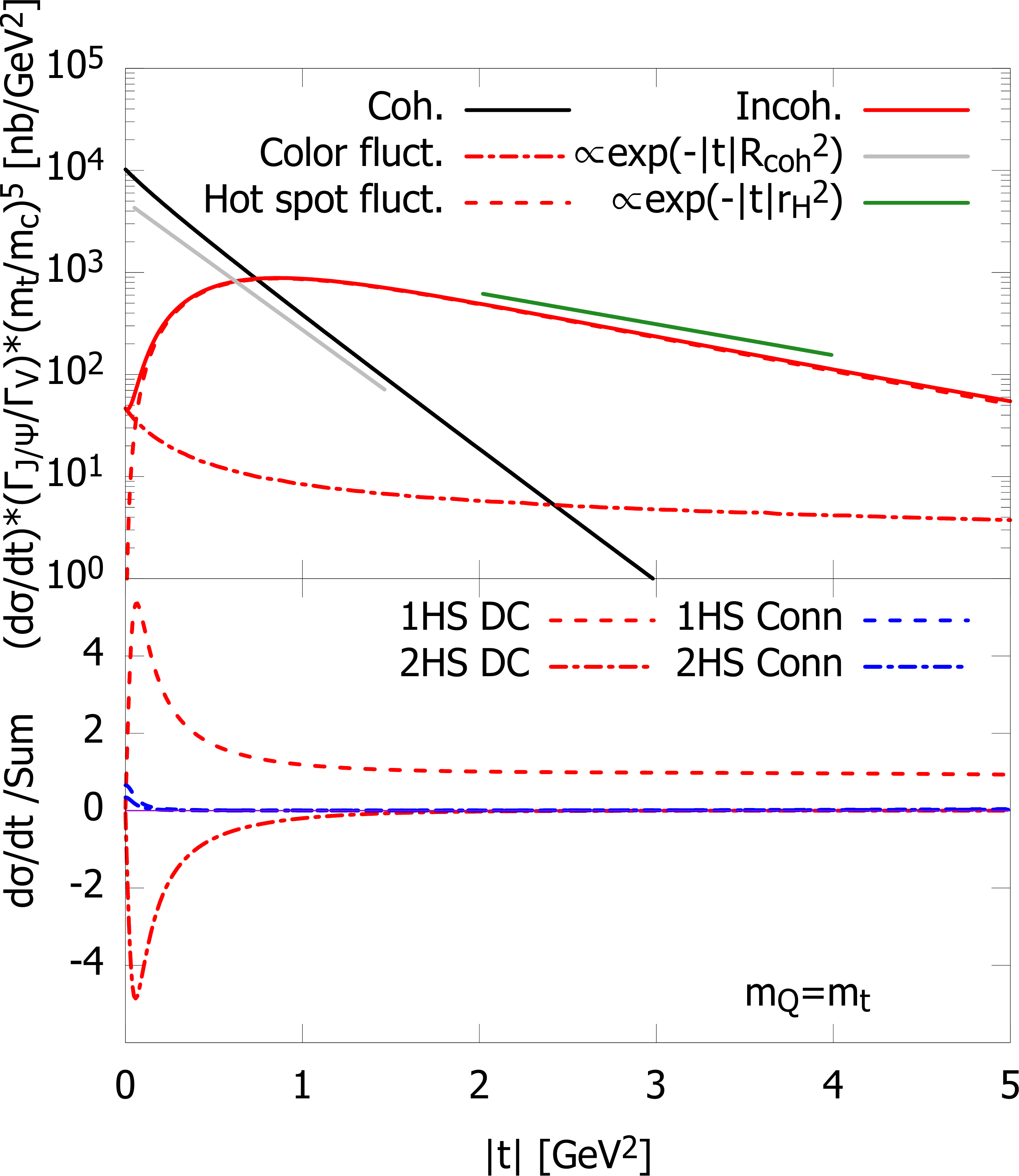}%
\includegraphics[height=1.03cm]{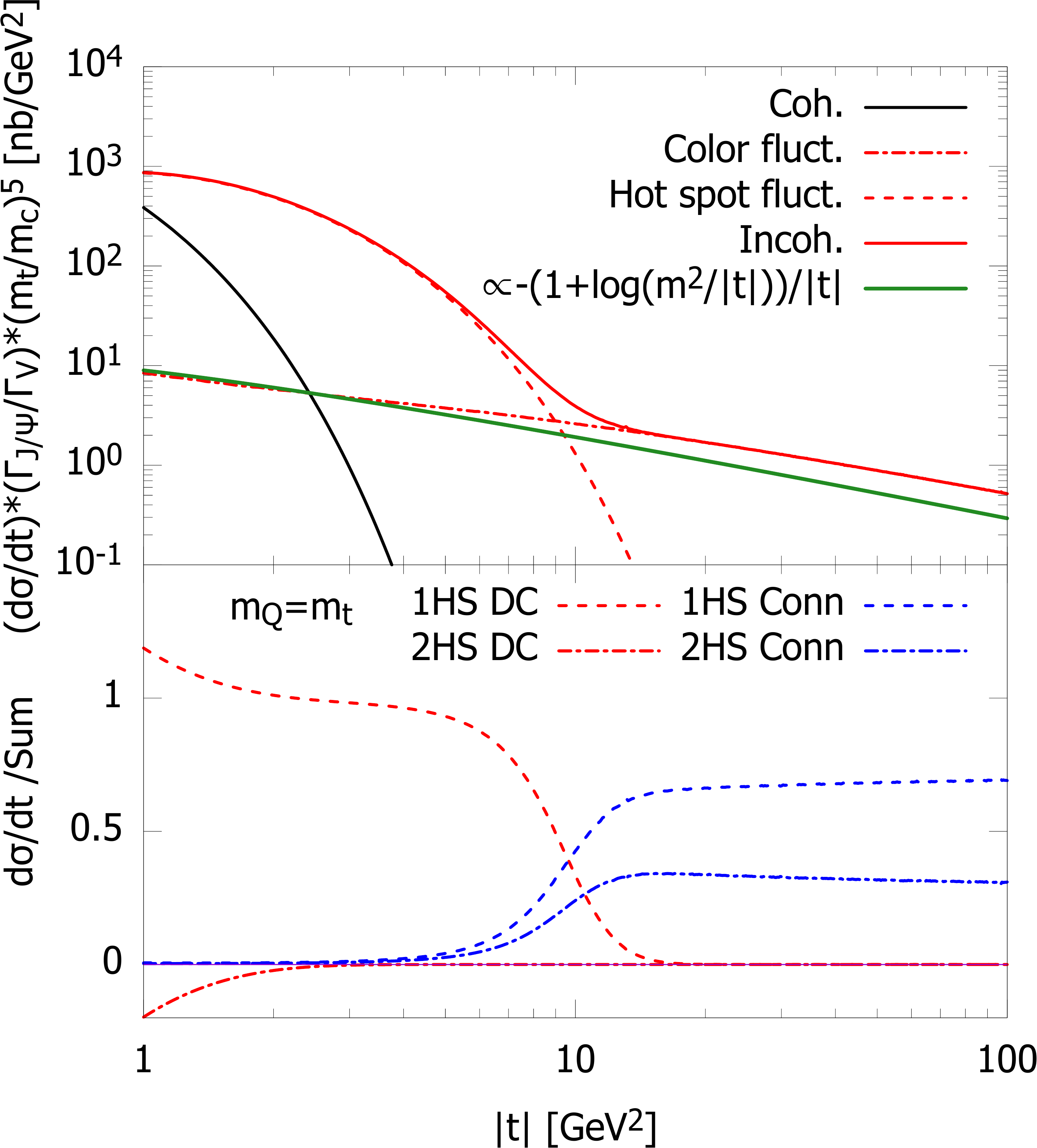}%
\includegraphics[height=1cm]{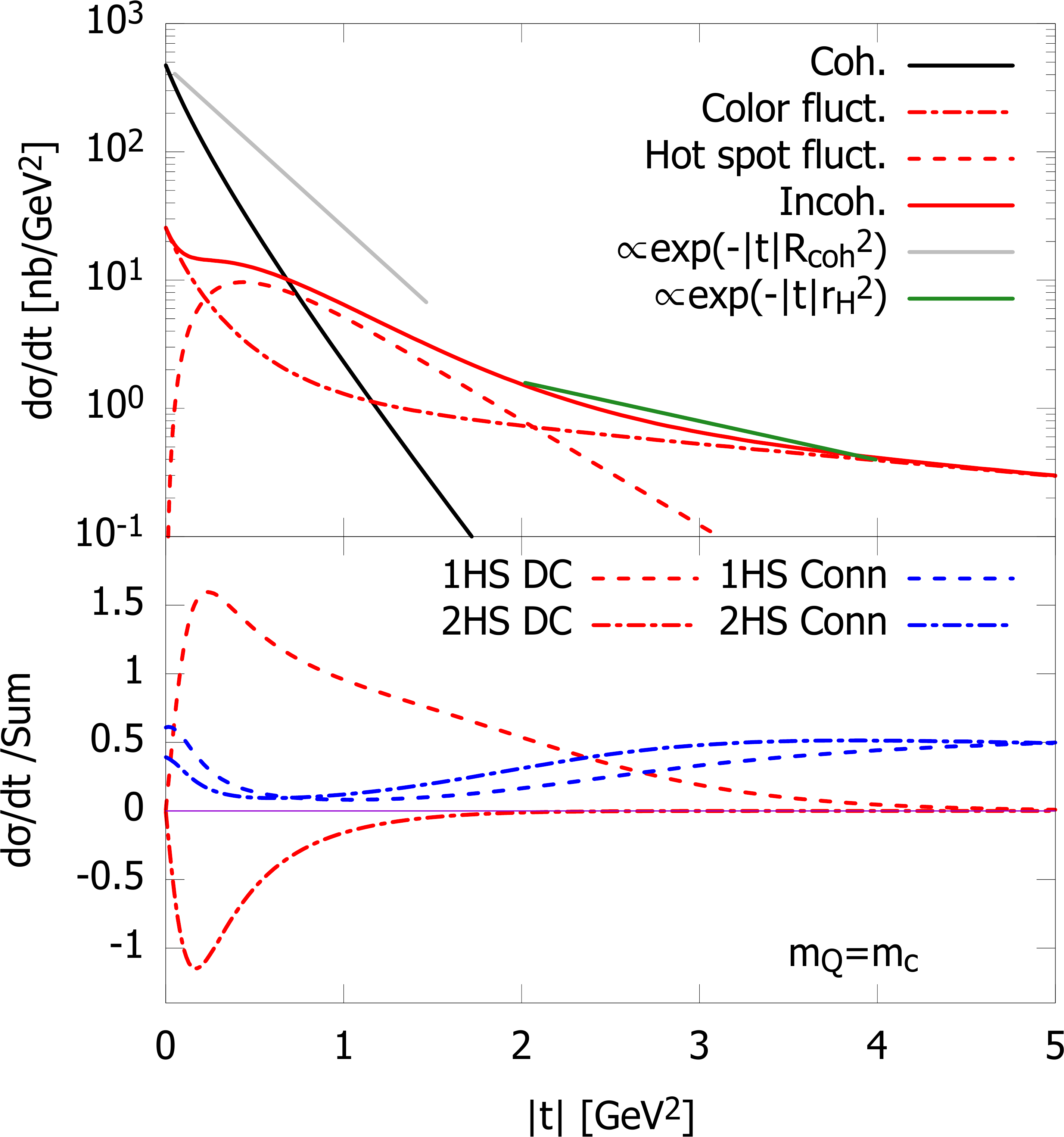}%
\includegraphics[height=1cm]{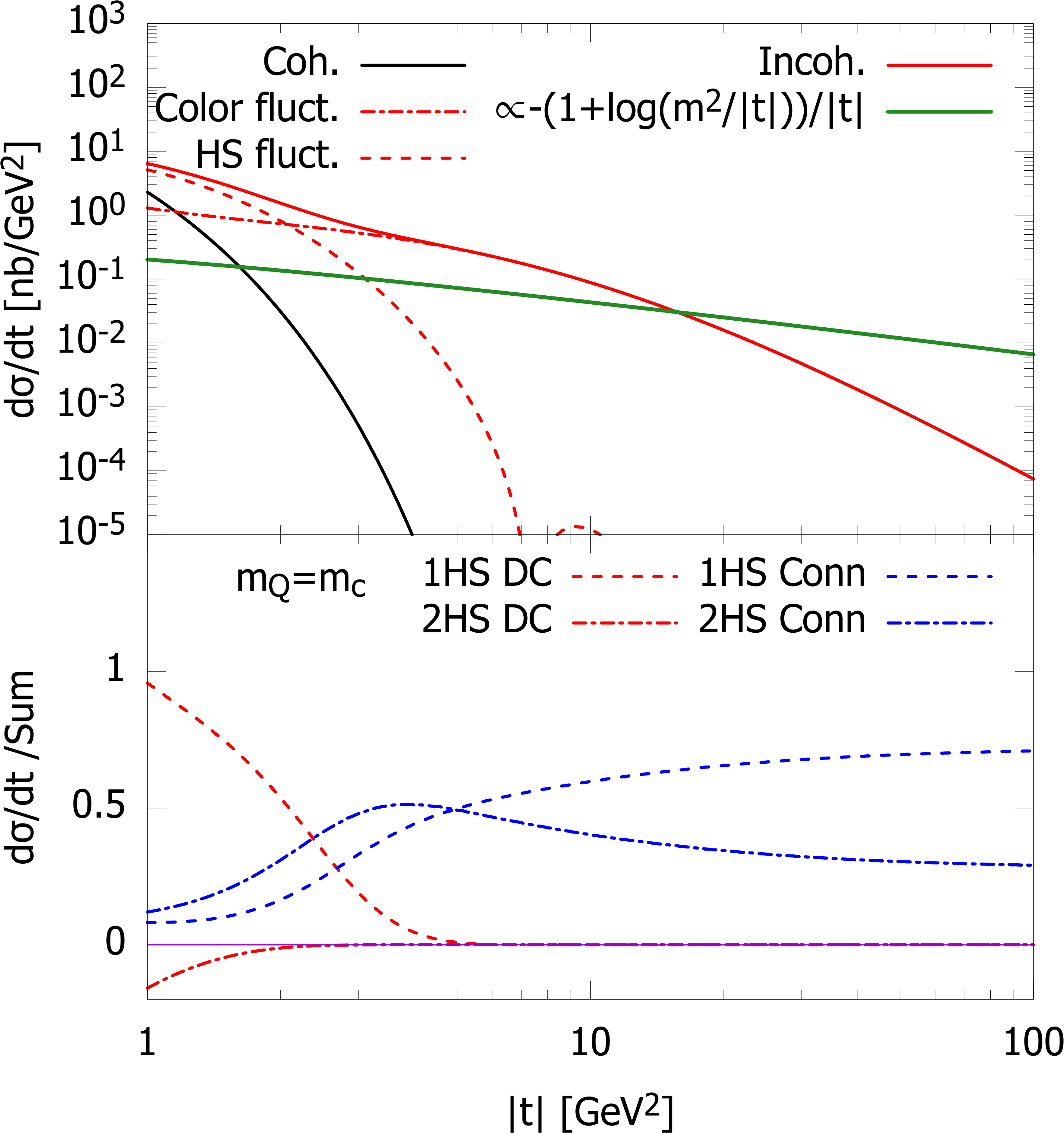}%
}
\caption{Different contributions to the coherent and incoherent spectrum for  large quark mass (two left plots) and charm (two right plots).}
\label{fig:charm}
\end{figure}

With  quite natural (compared to what is assumed in other studies) values for the parameters the $t$-dependence of the cross sections roughly agrees with HERA data. The only caveat here is that we have to articifially adjust up the normalization of the incoherent cross section. This indicates that we are missing a source of fluctuations that active up to very high $|t|$. In the model of Refs.~\cite{Mantysaari:2016ykx,Mantysaari:2016jaz} this is provided by ``$\qs$ fluctuations''; in our framework a more natural assumption would be to have $N_q$ fluctuate. The cross sections, split into contributions from different sources of fluctuations, are shown in Fig.~\ref{fig:charm}  for a very heavy quark mass and  for charm. One can distinguish some characteristic behaviors in different ranges in $t$. For very heavy quarks the coherent cross section dominates the small $|t$ behavior, where its slope is not quite $R_c^2 \equiv \frac{N_q-1}{N_q}R^2+r_H^2$ as one would expect from the color charges, but broader by $\sim 1/m$, since the amplitude measures the color field rather than the charge. The incoherent cross section is sensitive to the hot spot radius at 
intermediate values $|t|\sim 1/r_H^2$ and to color charge fluctuations at high $|t|$. For realistic charm quark masses, however, many of these interpretations break down. The slope of the coherent cross section is far from the target size. Although the intermediate $|t|\sim 1/r_H^2$ incoherent cross section slope is still related to $r_H$, the large-$|t|$ tails is far from what one would expect in the ideal ($m_Q\to \infty$) limit.

\section{Conclusions}
In conclusion, there is an emerging consensus that a ``gluonic hot spot'' structure in the nucleon is important for many high energy processes, for proton-nucleus and even nucleus-nucleus collisions. This structure can be probed by incoherent exclusive vector mesons in DIS and in ultraperipheral collisions. We developed a simple hot spot model to  average over fluctuations analytically, and used it to calculate eccentricities in proton-nucleus collisions, which turn out to be very sensitive to the hot spot structure. The parameters of such models can be extracted from exclusive vector meson production data, but we showed that for realistic quark masses the intepretation of the $t$-dependence of the vector meson cross section is far from what one would assume based on the $m_Q\to \infty$ limit.

\begin{spacing}{0.8}
 \paragraph{Acknowledgements}
{\footnotesize \setlength{\lineskip}{-3pt} S.D. acknowledges the support of the Vilho, Yrjö and Kalle Väisälä Foundation.
S.S acknowledges support by the Deutsche Forschungsgemeinschaft (DFG, German Research Foundation) through the CRC-TR 211 'Strong-interaction matter under extreme conditions'– project number 315477589 – TRR 211.
S.D. and T.L have  been supported by the Academy of Finland, by the Centre of Excellence in Quark Matter (project 346324) and project 321840. This work has also been supported under the European Union’s Horizon 2020 research and innovation programme by the STRONG-2020 project (grant agreement No 824093) and by the European Research Council, grant agreements ERC-2015-CoG-681707 and ERC-2018-AdG-835105. The content of this article does not reflect the official opinion of the European Union and responsibility for the information and views expressed therein lies entirely with the authors. 
}
\end{spacing}

\bibliography{qm23procs_lappi.bib}

\end{document}